\documentclass[11pt]{article}

\usepackage{array}
\usepackage{cite}
\usepackage[subrefformat=parens]{subcaption}
\usepackage{subdepth}
\usepackage{url}
\usepackage{fullpage} 
\usepackage{graphicx}
\usepackage{authblk}
\usepackage{amsmath}
\usepackage{amsfonts}
\usepackage{algorithm}
\usepackage{algorithmicx}
\usepackage{algpseudocode}
\usepackage{mathrsfs}

\usepackage[colorlinks=true,citecolor=blue,linkcolor=blue]{hyperref}
\usepackage[dvipsnames]{xcolor}

\renewcommand{\arraystretch}{1.25}

\title{Transport of Event Equation: Phase Retrieval from Defocus Events}

\author[1]{Kaito Hori}
\author[1\footnote{ctsutake@nagoya-u.jp}]{Chihiro Tsutake}
\author[1]{Keita Takahashi}
\author[1]{Toshiaki Fujii}

\affil[1]{
Department of Information and Communication Engineering, 
Nagoya University, 
Furo-cho, 
Chikusa-ku, 
Nagoya, 
464-8603, 
Japan}

\date{\empty}

\begin{document}

\maketitle
\vspace{-10mm}

\begin{abstract}
To time-efficiently and stably acquire the intensity information for phase retrieval under a coherent illumination, we leverage an event-based vision sensor~(EVS) that can detect changes in logarithmic intensity at the pixel level with a wide dynamic range.
In our optical system, we translate the EVS along the optical axis, where the EVS records the intensity changes induced by defocus as events.
To recover phase distributions, we formulate a partial differential equation, referred to as the \textit{transport of event equation}, which presents a linear relationship between the defocus events and the phase distribution.
We demonstrate through experiments that the EVS is more advantageous than the conventional image sensor for rapidly and stably detecting the intensity information, defocus events, which enables accurate phase retrieval, particularly under low-lighting conditions.
\end{abstract}


\section{Introduction}
\label{s1}
Phase retrieval~\cite{Shechtman:15} is a fundamental problem in optics in which the aim is to recover the phase distribution of an optical wavefront from intensity-only measurements under a coherent illumination.
In adaptive optics~\cite{Tyson:22}, real-time phase retrieval is essential for rapidly unveiling wavefront aberrations in the optical system.
To address this challenge, various solution methods have been proposed, which can be classified into two classes: analytical methods~\cite{Burge:76,Wood:81,Teague:83,Streibl:84} and iterative methods~\cite{Gerchberg:72,Fienup:82,Candes:15,Goldstein:18}.
Among these methods, a practical solution is offered by the transport of intensity equation~(TIE)~\cite{Teague:83,Streibl:84}, which describes the linear relationship between the phase distribution and the axial intensity derivative\footnote{The rate of intensity changes along the optical direction.}.
Unlike iterative methods, the TIE can be instantly solved using fast Fourier transform algorithms without the iteration~\cite{Allen:01}.
This characteristic is convenient for real-time phase retrieval.

With the aim of phase retrieval based on the TIE, we investigate methods for measuring the axial intensity derivative time-efficiently and stably.
The most straightforward approach is to compute the difference between two intensity images captured at different defocus planes~\cite{Teague:83,Streibl:84}.
However, capturing two distinct images requires waiting for the exposure time and can introduce temporal delays in adaptive optics systems.
Moreover, image sensors with a standard dynamic range\footnote{For example, $70.5$~dB of the Sony IMX265 image sensor.} are insufficient for stably capturing defocus images, especially under low-lighting conditions; typical examples of such conditions in adaptive optics include refractive error assessment~\cite{Guirao:03,Xu:15} and retinal imaging~\cite{Shirai:09}, where the power of a beam  incident on the human eyes must be minimized.
To accelerate the capturing of defocus images, single-shot methods have been proposed \cite{Waller:10a,Waller:10b,Angel:21,Hai:22}, where distinct defocus images are spatially multiplexed into a single image sensor.
However, single-shot methods inherently reduce the spatial resolution of each defocus image and remain unstable under low-lighting conditions.

\begin{figure}[!t]
\centering
\includegraphics[scale=1.1]{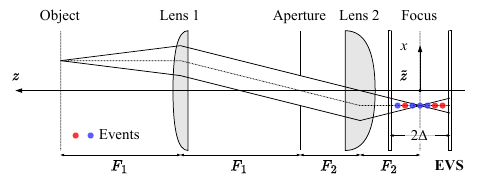}
\caption{Our optical system. $F_1$ and $F_2$ denote focal lengths of lenses~1 and 2, respectively. By telecentric characteristic~\cite{Kingslake:10}, $F_2/F_1$-scale object is projected onto focus plane. Event vision sensor~(EVS) is movable along optical axis $z$. We recover phase distribution on focus plane from events triggered by defocus.}
\label{fig:setup}
\end{figure}

To address these limitations, we leverage an event-based vision sensor (EVS)~\cite{Lichtsteiner:08}, which can detect changes in logarithmic intensity at the pixel level with a wide dynamic range\footnote{For example, $120$~dB of the Sony IMX636 EVS.}.
Figure~\ref{fig:setup} illustrates a schematic diagram of our optical system.
We translate the EVS along the optical axis, where the EVS records the intensity changes induced by defocus as events.
Then, we obtain the axial \textbf{log-}intensity derivative by integrating the recorded events pixel-by-pixel.
We finally recover a phase distribution on the basis of \textit{transport of event equation~(TEE)}, which presents a linear relationship between the axial log-intensity derivative~(defocus events) and the phase distribution.
Unlike previous methods~\cite{Teague:83,Streibl:84,Waller:10a,Waller:10b,Angel:21,Hai:22}, our method can avoid directly acquiring defocus images and thus does not require waiting for the exposure.
Moreover, the wide dynamic range of the EVS enables the axial intensity derivative to be stably detected under low-lighting conditions.
The effectiveness of these characteristics is demonstrated through computer simulations and optical experiments.
\vspace{3mm}

\noindent\textbf{Remark~1:} 
The TIE is capable of recovering continuous (unwrapped) phase distributions.
As our TEE will be derived from the TIE, it inherits the ability of phase unwrapping.

\section{Preliminaries}
\label{s2}
As shown in Fig.~\ref{fig:setup}, without loss of generality, we consider 1D spatial planes in the object and image spaces; all the discussions hereafter can be extended to a 2D setup by increasing the dimensions of the planes.
Notations and definitions are summarized as follows.
Let $x$ and $z$ be the spatial and depth coordinates, respectively.
We represent the intensity and phase values on $(x,z)$ as $I_{x,z}$ and $\phi_{x,z}$, respectively.
Let $\Delta$ be a distance between the focus and defocus planes.
Let $\tilde{z}$ represent the depth position of the focus plane.
We denote the intensity values of two defocus images captured at $(x,\tilde{z}\pm\Delta)$ by $I_{x,\tilde{z}\pm\Delta}$.

\section{Transport of intensity equation}
\label{s3}
In this section, we aim to recover the phase distribution $\phi_{x,\tilde{z}}$ on the focus plane from the defocus-image intensities since image sensors can measure the intensity only.
Throughout this paper, we assume that the intensity distribution at the focus plane is constant\footnote{Typical examples of such objects/images include the lens (refractive index distribution) of the human eye~\cite{Garner:98}.}, i.e., $I_{x,\tilde{z}} = I$.
Under this assumption, the TIE can be formulated as the following elliptic partial differential equation~(a.k.a. Poisson equation)~\cite{Teague:83,Streibl:84}:
\begin{equation}
\label{eq:tie}
\frac{\partial^2}{\partial x^2}\phi_{x,\tilde{z}}=-\frac{k}{I}\frac{\partial}{\partial z}I_{x,\tilde{z}},
\end{equation}
where $k$ is the wavenumber and $\partial I_{x,\tilde{z}}/\partial z$ is referred to as the axial linear-intensity derivative.
The closed-form solution of the TIE is as follows~\cite{Allen:01}:
\begin{equation}
\label{eq:phi_tie}
\phi_{x,\tilde{z}}=\mathscr{F}^{-1}\bigg[\frac{1}{\omega^2+C}\mathscr{F}\bigg[-\frac{k}{I}\frac{\partial}{\partial z}I_{x,\tilde{z}}\bigg]\bigg],
\end{equation}
where $\mathscr{F}$ is the Fourier transform operator along the spatial axis $x$, $\omega$ is the spatial frequency, and $C$ is a constant for suppressing noise amplification~\cite{Wiener:49}.
Note that $\omega^2$ and $\omega^{-2}$ are the forward and inverse transfer functions of $\partial^2/\partial x^2$, respectively.

The axial linear-intensity derivative can be computed using the defocus images $I_{x,\tilde{z}\pm\Delta}$ as follows:
\begin{equation}
\label{eq:linear}
\frac{\partial}{\partial z}I_{x,\tilde{z}}\simeq\frac{I_{x,\tilde{z}+\Delta} - I_{x,\tilde{z}-\Delta}}{2\Delta}.
\end{equation}
However, capturing $I_{x,\tilde{z}\pm\Delta}$ is time-consuming due to exposure times and unstable under low-lighting conditions. 
The instability primarily arises from the following four processes, which are often assumed in the image restoration literature~\cite{Foi:08,Luisier:11}.
\begin{enumerate}
\item \textit{Poisson process} reproduces shot noise due to the stochastic nature of photon arrivals.
\item \textit{Gaussian process} models electronic readout noise. We assume a zero-mean Gaussian process with the standard deviation $\sigma_\mathrm{TIE}$.
\item \textit{Clipping} limits intensity values that exceed the sensor’s dynamic range prior to analog-to-digital conversion. It is particularly problematic under low-lighting conditions, leading to a loss of detail in dark regions. 
\item \textit{Quantization} reproduces rounding errors due to analog-to-digital conversion. Defocus changes smaller than the quantization step size are eliminated by this process.
\end{enumerate}

\section{Proposed method}
\label{s4}
Unlike conventional image sensors, the EVS records the time $t$, spatial coordinate $x$, and polarity of intensity changes (the red and blue points in Fig.~\ref{fig:setup}) during the translation.
We translate the EVS over a distance of $2\Delta$ during a time interval $2T$ while maintaining constant velocity.
The position of the EVS, denoted by $z_t \in [\tilde{z}-\Delta,\tilde{z}+\Delta]$, can be written as a function of $t$ as follows:
\begin{equation}
\label{eq:zt}
z_t = \frac{\Delta}{T}t+\tilde{z}- \Delta, \quad( 0 \leq t \leq 2T).
\end{equation}
An event at $(x,z_t)$, or equivalently $(x,t)$\footnote{For given $\tilde{z}$ and $z_t$, $t$ can be determined from \eqref{eq:zt}.}, is triggered when the intensity change (in log scale) exceeds a contrast threshold $\mu$~\cite{Lichtsteiner:08}, i.e.,
\begin{equation}
\label{eq:event}
|\log I_{x,z_t}-\log I_{x,z_{t_0}}| > \mu,
\end{equation}
where $z_{t_0}$ is the depth when the previous event was recorded.
We integrate events along $t$ for each $x$; the resulting data form a single plane (as will be illustrated in Figs.~\ref{fig:sim_oa} and \ref{fig:sim_ob}) and are denoted by $E_{x,\tilde{z}-\Delta\rightarrow \tilde{z}+\Delta}$.
Suppose that there is no noise in the recorded events.
Under this assumption, $E_{x,\tilde{z}-\Delta\rightarrow \tilde{z}+\Delta} $ and \eqref{eq:event} have the following relationship.
\begin{equation}
\label{eq:acc}
E_{x,\tilde{z}-\Delta\rightarrow \tilde{z}+\Delta}=\frac{\log I_{x,\tilde{z}+\Delta} - \log I_{x,\tilde{z}-\Delta}}{\mu}
\end{equation}
Equation~\eqref{eq:acc} indicates that the EVS captures intensity changes induced by defocus as relative rates in the logarithmic domain, rather than the differences in the linear domain~\eqref{eq:linear}.
As a result, the sensitivity of the EVS remains consistent regardless of the intensity scale.
This characteristic enables the EVS to stably detect fine defocus changes even under low-lighting as well as high lighting conditions\footnote{In other words, the EVS operates over a wide dynamic range.}.
Moreover, event acquisition is more time-efficient than capturing defocus images with exposure times.

We aim to reconstruct $\phi_{x,\tilde{z}}$ from $E_{x,\tilde{z}-\Delta\rightarrow \tilde{z}+\Delta}$.
We use the following elementary relationship:
\begin{equation}
\label{eq:diff}
\frac{1}{I}\frac{\partial}{\partial z} I_{x,\tilde{z}}=\frac{\partial}{\partial z} \log I_{x,\tilde{z}}\simeq \mu\frac{E_{x,\tilde{z}-\Delta\rightarrow \tilde{z}+\Delta}}{2\Delta}.
\end{equation}
Substituting \eqref{eq:diff} into \eqref{eq:tie}, we obtain the following Poisson equation, referred to as the TEE:
\begin{equation}
\frac{\partial^2}{\partial x^2} \phi_{x,\tilde{z}}=-k\frac{\partial}{\partial z} \log I_{x,\tilde{z}}.
\label{eq:tee}
\end{equation}
The solution can be obtained by replacing $I^{-1}\partial I_{x,\tilde{z}}/\partial z$ in \eqref{eq:phi_tie} with $\partial \log I_{x,\tilde{z}}/\partial z$.
\vspace{3mm}

\noindent\textbf{Remark~2:} The TEE implies that the recorded events contain the curvature information, i.e., the second derivative, of the phase distribution.
Moreover, the phase distribution can be recovered (up to a global phase ambiguity\footnote{As can be seen from \eqref{eq:phi_tie}, both the TIE and TEE cannot recover the direct current~($\omega=0$).}) from the recorded events.
We believe that these characteristics provide novel insights into the fundamental capabilities of the EVS.

\vspace{3mm}
\noindent\textbf{Remark~3:} In contrast to the TIE, the TEE does not contain the absolute intensity value $I$ in the right-hand side to obtain correctly scaled phase distributions.
In other words, our method operates without measuring $I$, which is a key advantage.
\vspace{3mm}

\noindent\textbf{Remark~4:}
Our method has potential applications in display systems.
In \cite{Peng:20,Yamamoto:21}, 
optical aberrations in display systems, such as refractive errors of the eye, 
are measured and fed back to the systems to improve image quality.
Our TEE can be incorporated into these approaches as a method for measuring optical aberrations.
\vspace{3mm}

\begin{figure}[!t]
\centering
\includegraphics[scale=1.1]{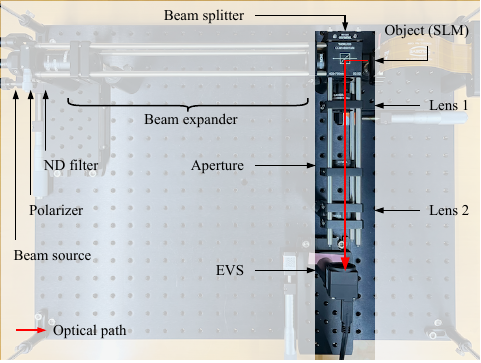}
\caption{Hardware implementation of our optical system. EVS is movable along optical axis (red arrow).}
\label{fig:impl}
\end{figure}

As mentioned in \cite{Hu:21}, observed events are actually degraded through Poisson, Gaussian, and quantization processes as follows.
\begin{enumerate}
\item \textit{Poisson process} reproduces shot noise and is applied to the intensities $I_{x,\tilde{z}\pm \Delta}$ in \eqref{eq:acc}.
\item \textit{Gaussian process} models fluctuation of the contrast threshold $\mu$. The zero-mean Gaussian noise with the standard deviation $\sigma_\mathrm{TEE}$ is added to $\mu$. 
\item \textit{Quantization} introduces a rounding error in $E_{x, \tilde{z}-\Delta\rightarrow\tilde{z}+\Delta}$ since the polarities and their integrals must be integers.
\end{enumerate}
To suppress noise amplification, we carefully choose the constant $C$ from the candidates $10^{-2}$, $10^{-3}$, $\cdots$, $10^{-13}$ depending on the noise level.

\begin{table*}[!t]
\caption{Weights of Zernike polynomials (blank: 0). Symbol $i$ represents OSA standard index~\cite{Foi:22} and corresponds to optical aberration, e.g., horizontal tilt ($i=2$), defocus ($i=4$), and vertical coma ($i=7$).}
\label{tab:zernike}
\centering
\footnotesize
\renewcommand{\arraystretch}{1.2}
\begin{tabular}{ccccccccccccccc}\hline
$i$              &   2&   3&   4&   5&   7&   8&  12&  13&  14&  16&  17&  19&  24&  27\\\hline
\textit{Phase~0} &0.08&    &0.10&0.05&0.08&    &    &    &    &    &    &    &    &    \\
\textit{Phase~1} &    &    &0.08&    &    &0.04&0.10&    &    &    &    &0.06&    &0.07\\
\textit{Phase~2} &    &0.05&    &    &    &    &    &0.11&    &    &0.06&    &    &    \\
\textit{Phase~3} &    &    &    &    &    &    &    &    &0.03&0.10&    &    &0.08&    \\\hline
\end{tabular}
\vspace{4mm}
\end{table*}

Figure~\ref{fig:impl} shows the hardware implementation of our optical system.
We employed the phase-only spatial light modulator~(SLM) as the object with uniform intensity.
Target phase distributions to be retrieved are displayed on this SLM.
To simulate low-lighting conditions, we placed an ND filter that reduces the intensity of the incident beam.
Parameters of optical elements will be detailed in Section~\ref{s5ss2}.

\section{Results}
\label{s5}
\subsection{Computer simulation}
\label{s5ss1}
To verify the effectiveness of our method without any optical aberrations, we conducted a computer simulation of the optical system illustrated in Fig.~\ref{fig:setup}.
We used weighted sums of Zernike polynomials~\cite{Foi:22} as the target phase distributions for reconstruction.
The weight values are listed in Table~\ref{tab:zernike}.
The pupil diameter for the Zernike polynomials was set to $3.2$~mm.
The wavelength of the illumination beam was $635$~nm.
The translation distance $2\Delta$ was $40$~mm.
The wave propagation between $I_{x,\tilde{z}}$ and $I_{x,\tilde{z}\pm \Delta}$ was computed using the Fresnel diffraction integral~\cite{Kelly:14}.
The intensity level $I_{x,\tilde{z}}=I$ was varied from $0.1$ to $100$ to control the strength of shot noise.
We assumed that the EVS has the minimum measurable intensity value of $I = 0.1$; intensity values less than $0.1$ were rounded up to $0.1$.
A phase distribution was placed on the focus plane and retrieved from the axial log-intensity derivative $E_{x,\tilde{z}-\Delta\rightarrow\tilde{z}+\Delta}$.
The contrast threshold was $\mu=0.1$.
We used $\sigma_\mathrm{TEE} = 0.03$ to simulate Gaussian noise for $\mu$.
The pixel pitch and resolution of the EVS were $6.4$~\textmu m and $539 \times 539$, respectively.
For quantitative evaluation,  we calculated the root-mean-square error~(RMSE) between the original and retrieved phase values.

\begin{figure}[!t]
\centering
\includegraphics[scale=0.24]{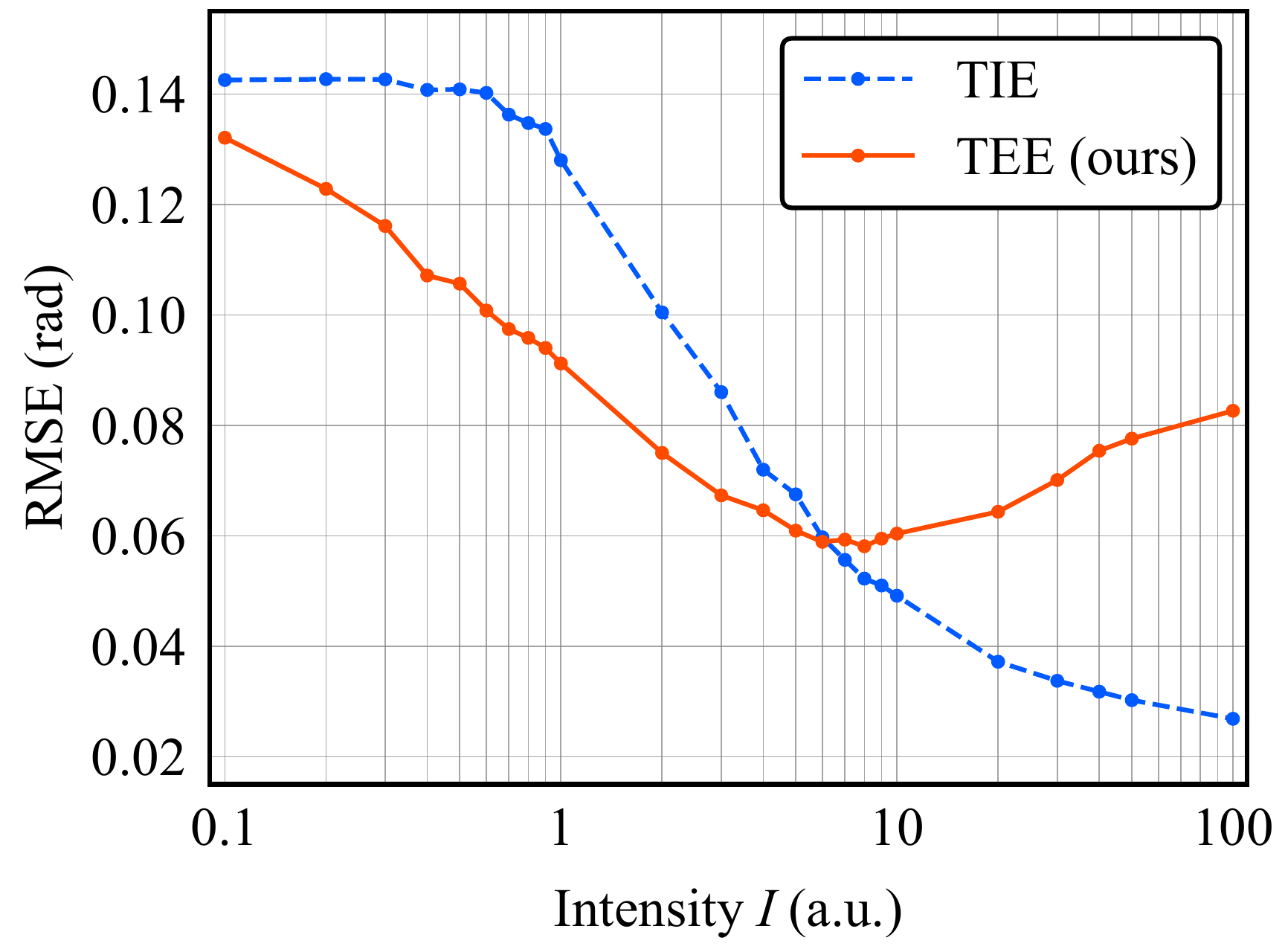}
\caption{RMSE values against intensity levels $I$.}
\label{fig:rmse}
\end{figure}

\begin{figure}[!t]
\centering
\begin{subfigure}{0.45\textwidth}
\centering
\includegraphics[scale=1.1]{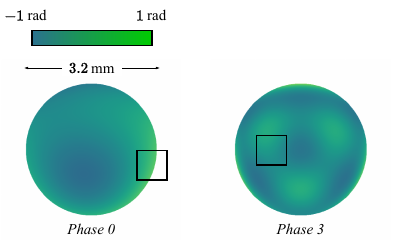}
\caption{Original}
\end{subfigure}
\begin{subfigure}{0.45\textwidth}
\centering
\includegraphics[scale=1.1]{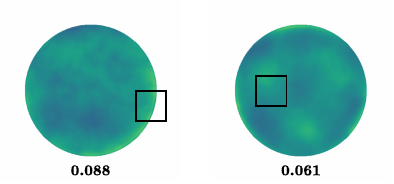}
\caption{TEE (pixel pitch: 6.4~\textmu m)}
\end{subfigure}\\
\centering
\begin{subfigure}{0.45\textwidth}
\centering
\includegraphics[scale=1.1]{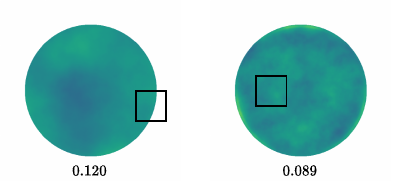}
\caption{TIE (pixel pitch: 6.4~\textmu m)}
\end{subfigure}
\begin{subfigure}{0.45\textwidth}
\centering
\includegraphics[scale=1]{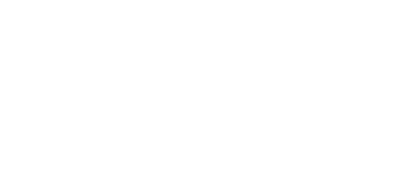}
\end{subfigure}
\caption{Retrieved phases w/ RMSE (simulation, $I=2$). Black frames in (b) and (c) highlight regions with small and large errors, respectively.}
\label{fig:sim_p}
\end{figure}

\begin{figure}[!t]
\centering
\begin{subfigure}{0.45\textwidth}
\hspace{-9mm}
\includegraphics[scale=0.24]{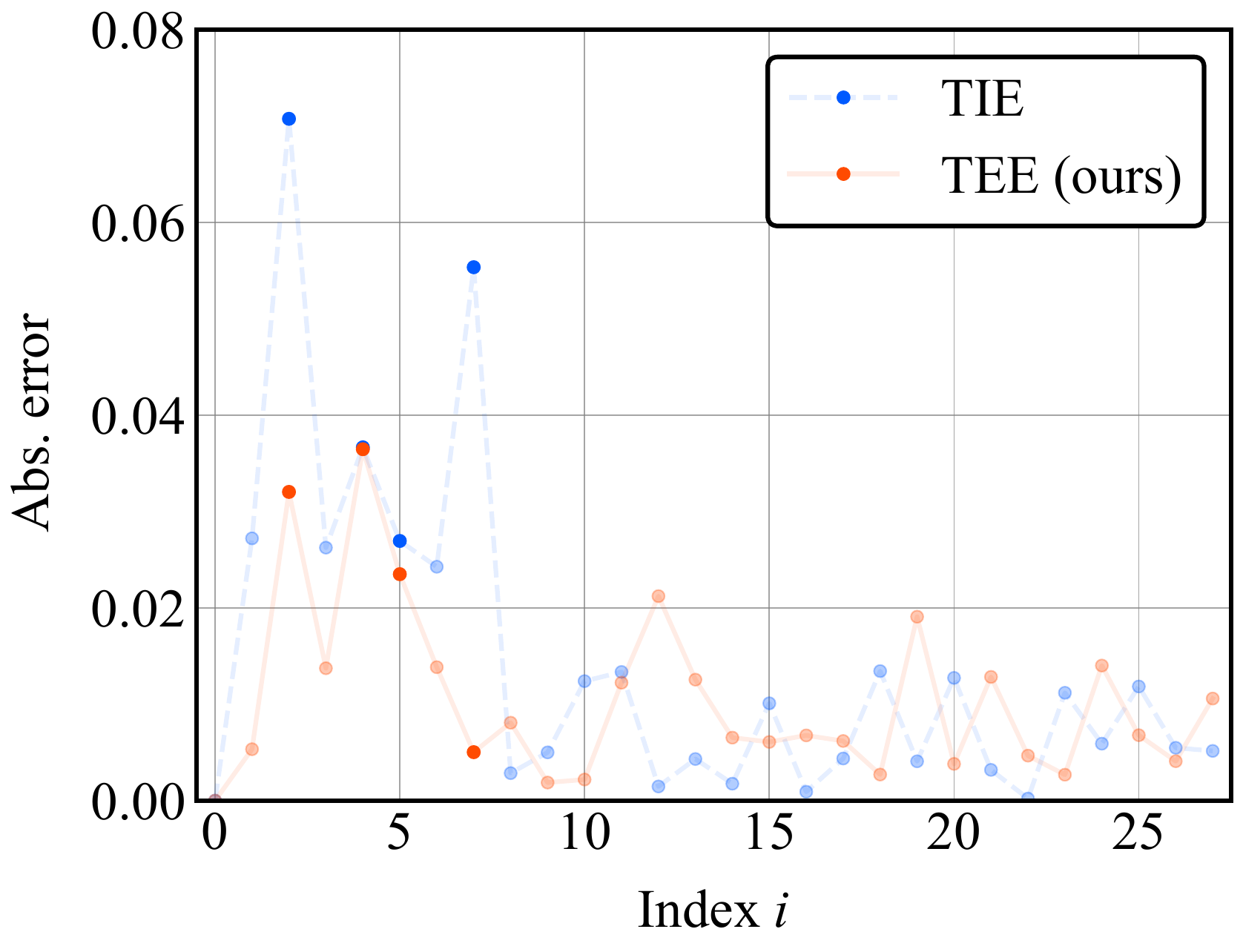}
\centering
\caption{\textit{Phase 0}}
\end{subfigure}
\begin{subfigure}{0.45\textwidth}
\hspace{-9mm}
\includegraphics[scale=0.24]{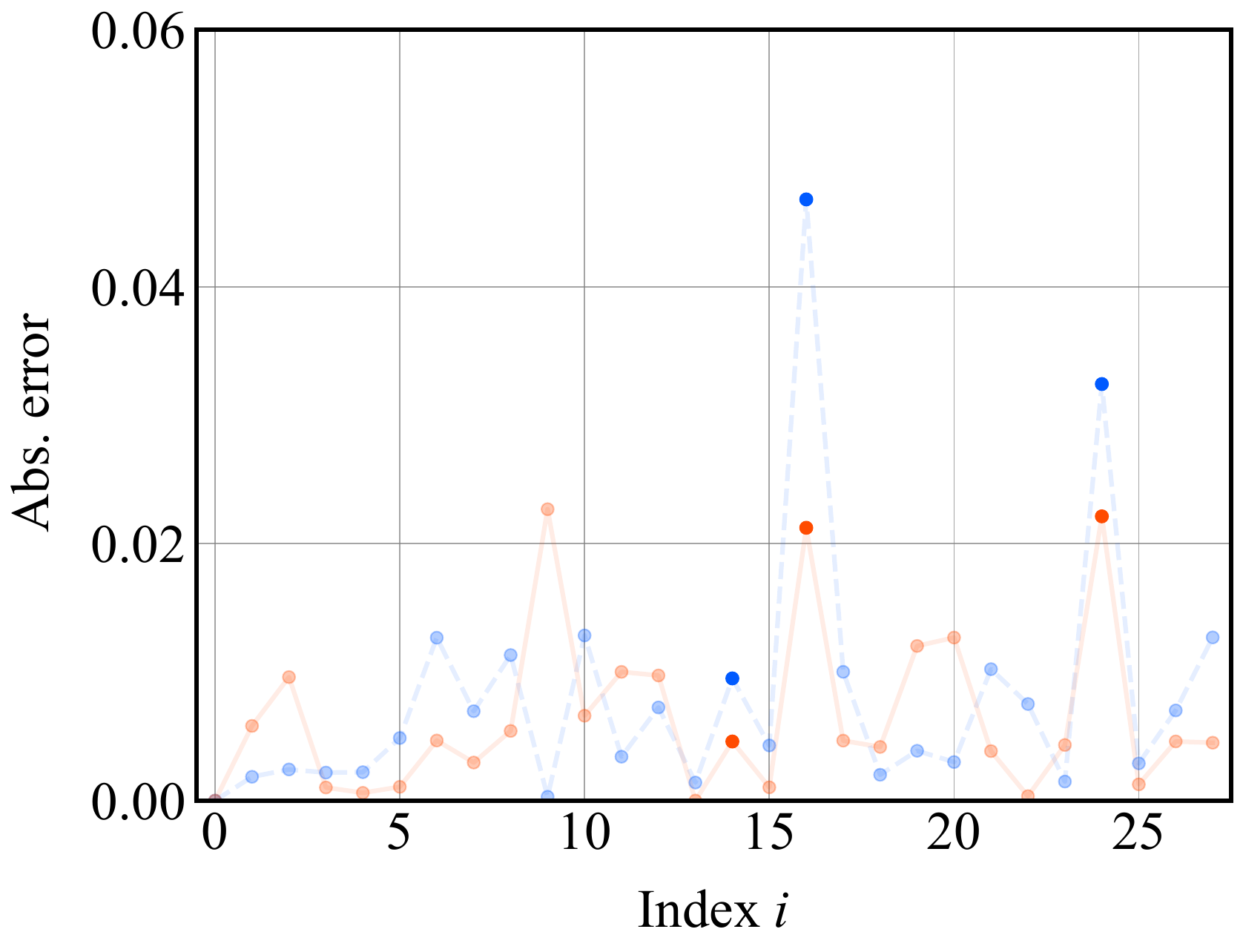}
\centering
\caption{\textit{Phase 3}}
\end{subfigure}
\caption{Reconstruction error of weights (simulation, $I=2$). Non-transparent points correspond to non-zero weights in Table~\ref{tab:zernike}.}
\label{fig:sim_e}
\end{figure}

\begin{figure}[!t]
\centering
\vspace{-0.5mm}
\begin{subfigure}{0.45\textwidth}
\centering
\includegraphics[scale=1.1]{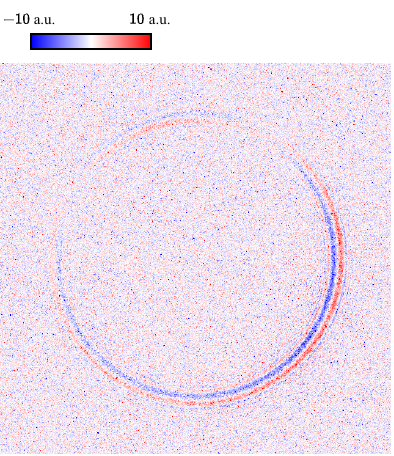}
\caption{$\log I_{x,\tilde{z}+\Delta}-\log I_{x,\tilde{z}-\Delta}$ in TEE}
\label{fig:sim_oa}
\end{subfigure}
\begin{subfigure}{0.45\textwidth}
\centering
\includegraphics[scale=1.1]{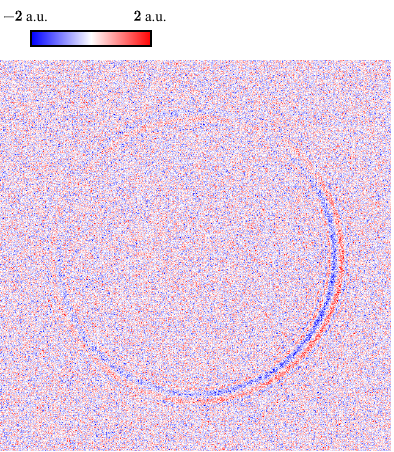}
\caption{$I_{x,\tilde{z}+\Delta}-I_{x,\tilde{z}-\Delta}$ in TIE}
\label{fig:sim_ob}
\end{subfigure}\\
\caption{Axial intensity derivative of \textit{Phase~0} (simulation, $I=2$).}
\label{fig:sim_o}
\end{figure}

\begin{figure}[!t]
\centering
\hspace{-8mm}
\includegraphics[scale=0.24]{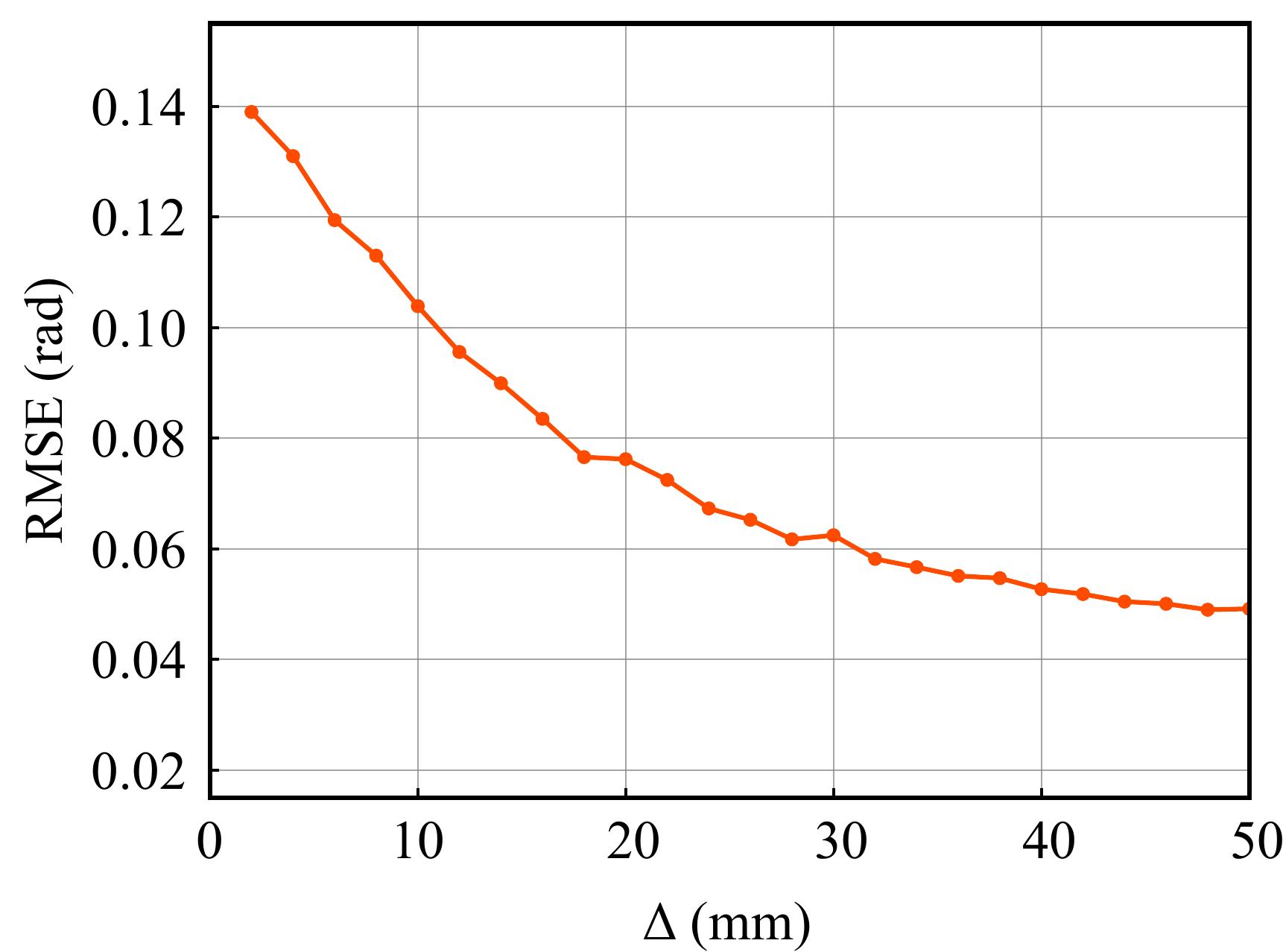}
\caption{RMSE values against translation distance $\Delta$ ($I=2$).}
\label{fig:delta}
\end{figure}

\begin{figure}[!t]
\centering
\begin{subfigure}{0.45\textwidth}
\centering
\includegraphics[scale=1.1]{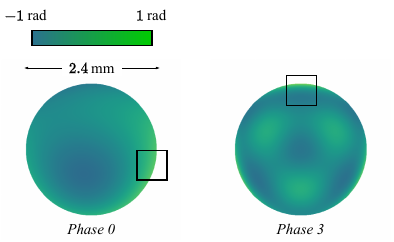}
\caption{Original}
\end{subfigure}
\begin{subfigure}{0.45\textwidth}
\centering
\includegraphics[scale=1]{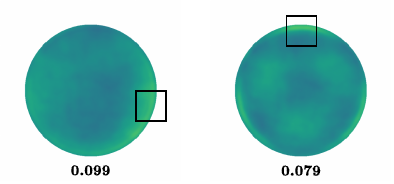}
\caption{TEE (pixel pitch: 4.86~\textmu m)}
\end{subfigure}\\
\begin{subfigure}{0.45\textwidth}
\centering
\includegraphics[scale=1.1]{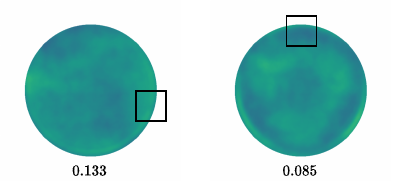}
\caption{TIE (pixel pitch: 3.45~\textmu m)}
\end{subfigure}
\begin{subfigure}{0.45\textwidth}
\centering
\includegraphics[scale=1]{dummy.pdf}
\end{subfigure}
\caption{Retrieved phases w/ RMSE (opt. expt.). Black frames in (b) and (c) highlight regions with small and large errors, respectively.}
\label{fig:opt_p}
\end{figure}

\begin{figure}[!t]
\centering
\begin{subfigure}{0.45\textwidth}
\hspace{-9mm}
\includegraphics[scale=0.24]{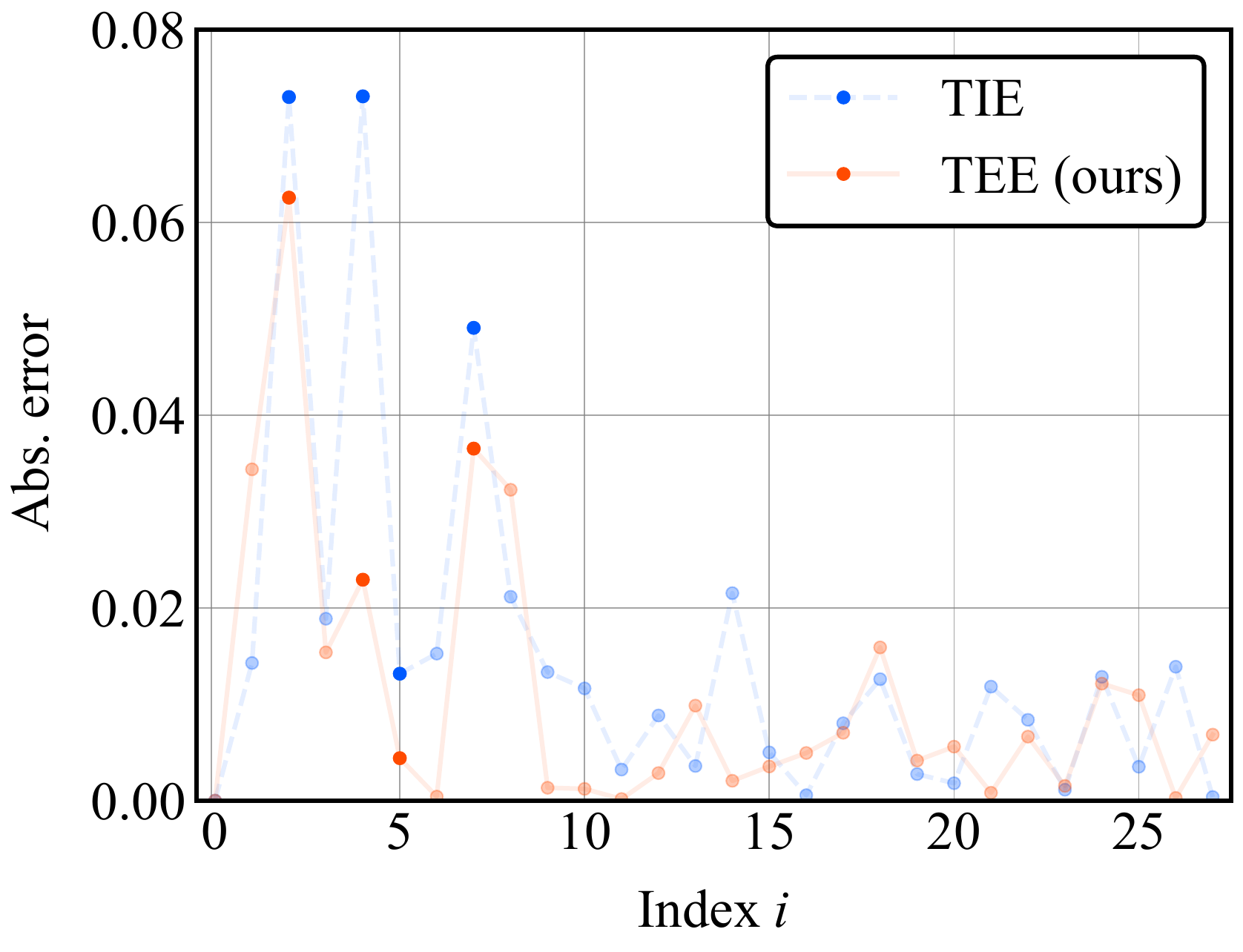}
\centering
\caption{\textit{Phase 0}}
\end{subfigure}
\begin{subfigure}{0.45\textwidth}
\hspace{-9mm}
\includegraphics[scale=0.24]{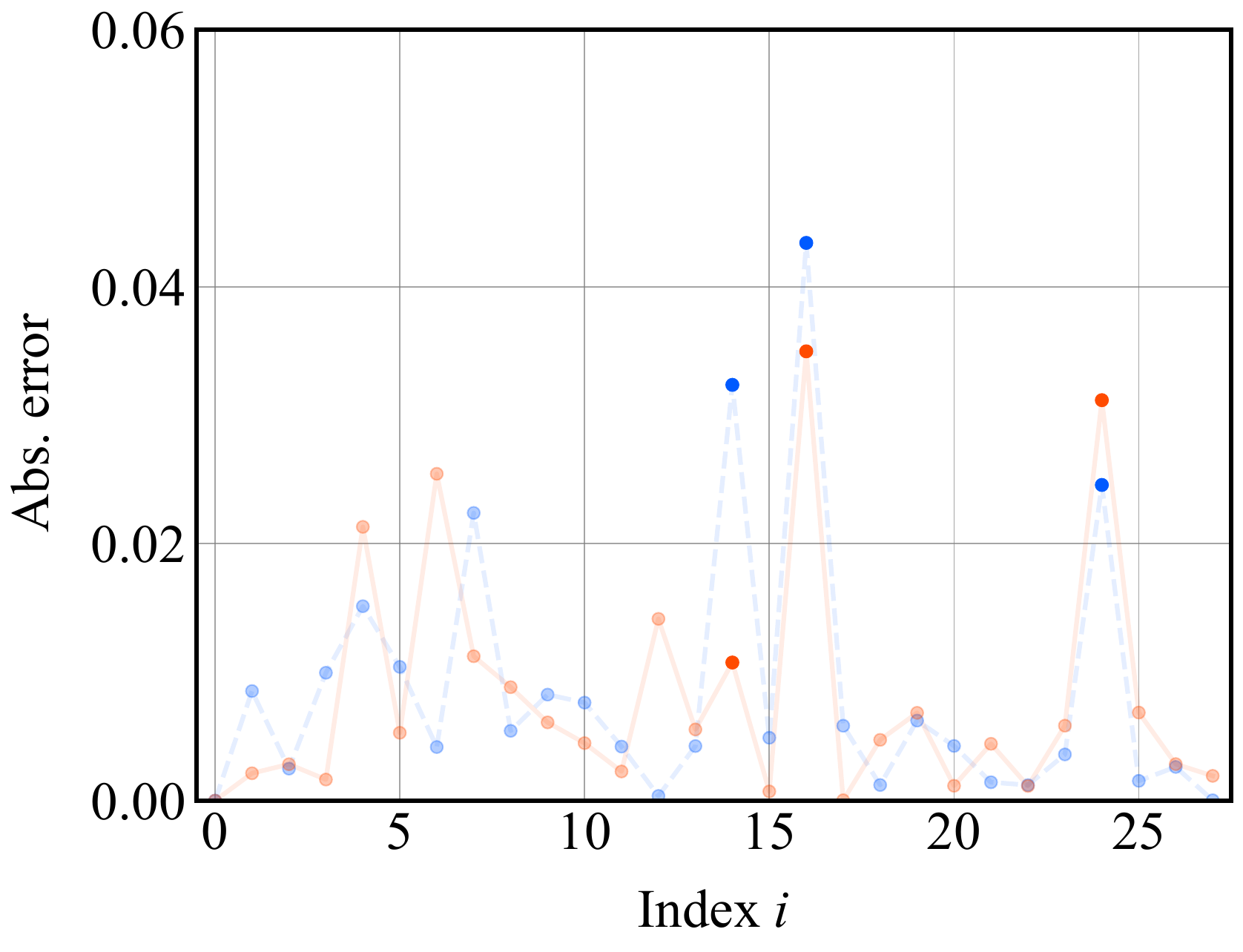}
\centering
\caption{\textit{Phase 3}}
\end{subfigure}
\caption{Reconstruction error of weights (opt. expt.). Non-transparent points correspond to non-zero weights in Table~\ref{tab:zernike}.}
\label{fig:opt_e}
\end{figure}

\begin{figure}[!t]
\centering
\begin{subfigure}{0.45\textwidth}
\centering
\vspace{-0.5mm}
\includegraphics[scale=1.1]{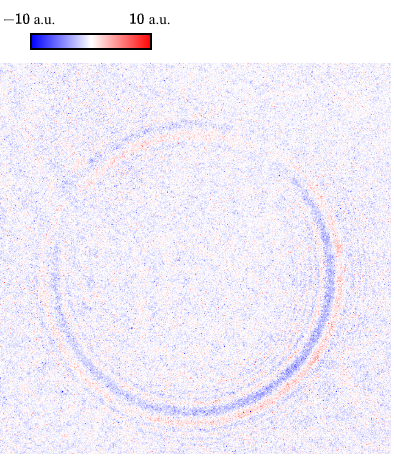}
\caption{$\log I_{x,\tilde{z}+\Delta}-\log I_{x,\tilde{z}-\Delta}$ in TEE}
\label{fig:opt_oa}
\end{subfigure}
\begin{subfigure}{0.45\textwidth}
\centering
\includegraphics[scale=1.1]{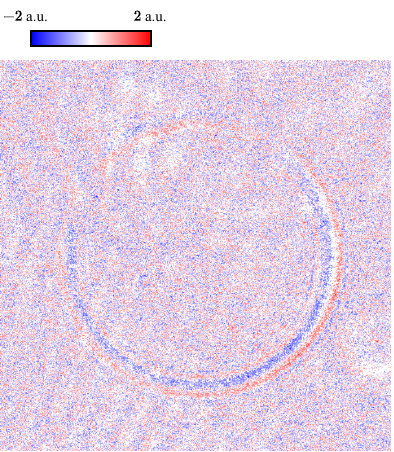}
\caption{$I_{x,\tilde{z}+\Delta}-I_{x,\tilde{z}-\Delta}$ in TIE}
\end{subfigure}\\
\caption{Axial intensity derivative of \textit{Phase~0} (opt. expt.).}
\label{fig:opt_o}
\end{figure}

For comparison, we captured the distinct two defocus images $I_{x,\tilde{z}\pm\Delta}$ and retrieved phase distributions using the TIE.
The image sensor was configured to have the same pixel pitch and resolution as the EVS.
The quantization step size was $1$.
We assumed that the image sensor has the minimum measurable intensity value of $I = 1$.
We used $\sigma_\mathrm{TIE} = 0.5$ to simulate Gaussian noise for defocus images.
The Neumann boundary condition was applied to the TEE and TIE.

Figure~\ref{fig:rmse} shows RMSE values against the intensity level $I$.
Our method consistently outperformed the TIE under low-intensity conditions, $I \leq 6$.
Figure~\ref{fig:sim_p} presents examples of the reconstructed phase distributions, where $I=2$.
We can visually confirm that our method recovered better phase distributions, as indicated by the square regions.
Figure~\ref{fig:sim_e} shows the absolute values of errors between the original and recovered weights of the Zernike polynomials.
Our method presented smaller errors than those by the TIE, particularly for the non-zero weights listed in Table~\ref{tab:zernike}.

We believe that these improvements are owing to the high dynamic range of the EVS, which allows stable detection even under low-lighting conditions.
Figure~\ref{fig:sim_o} shows axial intensity derivatives, where $I=2$.
Fine defocus changes around the edges can be recorded in the axial log-intensity derivative in Fig.~\ref{fig:sim_oa}.
On the other hand, minor defocus changes vanished in the axial linear-intensity derivative, as shown in Fig.~\ref{fig:sim_ob}, due to a narrow dynamic range and rough quantization step size of the image sensor.
We also reconstructed \textit{Phases~0 and 3} in the absence of Poisson and Gaussian noise to isolate their effects.
The TEE and TIE presented RMSE values of $\textbf{0.094}$ and $0.124$ on average, respectively.
These results support our statement. 
\vspace{3mm}

\noindent\textbf{Remark~5:}
We investigated the stability of the TEE with respect to variations of the translation distance.
We varied $2\Delta$ within the range $[4;100]$~mm.
We adopted $0.141$ as the acceptable RMSE value, 
which is the worst value obtained by the TEE through variations of $2\Delta$ at $I=0.1$.
When $I=2$ was fixed and $2\Delta$ varied, the RMSE values ranged from $0.049$ to $0.139$,
as shown in Fig.~\ref{fig:delta}.
This result indicates that the TEE stably presented RMSE values within the acceptable range across different $2\Delta$.
\vspace{3mm}

\subsection{Optical experiment}
\label{s5ss2}
We also conducted optical experiments using the system in Fig.~\ref{fig:impl}.
We used Holoeye phase-only SLM\footnote{Model: LETO-3, resolution: $1920 \times 1080$, pixel pitch: $6.4$~\textmu m.}.
\textit{Phases 0 and 3} in Table~\ref{tab:zernike} were displayed on this SLM.
The focal lengths $F_1$ and $F_2$ were $75$~mm and $50$~mm, respectively.
The wavelength and power of the beam were $635$~nm and $0.9$~mW, respectively.
The ND filter attenuates the beam power to $0.3$\%, resulting in an output power of $0.0027$~mW.
Prophesee EVK4 camera\footnote{Sensor: Sony IMX636 EVS, resolution: $1280\times720$, pixel pitch: 4.86~\textmu m, dynamic range: $120$~dB.} was placed on the defocus plane and translated over a distance $2\Delta = 25$~mm within $2T=1$~second.
All the events triggered during this period were considered as elements of the plane $E_{x,\tilde{z}-\Delta\rightarrow\tilde{z}+\Delta}$.
For comparison, as was done in the simulation, we captured defocus images and retrieved phases using the TIE, where we used Lucid Vision Lab TRI032S-CC camera\footnote{Sensor: Sony IMX265 image sensor, resolution: $2048\times1536$, pixel pitch: 3.45~\textmu m, dynamic range: $70.5$~dB.}.
An exposure time of $3000$~microseconds was used for capturing defocus images.
Note that the EVS operates without exposure time.

Figures~\ref{fig:opt_p}, \ref{fig:opt_e}, and \ref{fig:opt_o} present reconstructed phase distributions, errors of recovered weights, and axial intensity derivatives, respectively.
Similar results in the simulation were observed in the optical experiments as well.
These results demonstrate the advantages of our method in terms of acquiring the axial intensity derivative and achieving accurate phase retrieval under low-lighting conditions.
\vspace{3mm}

\noindent\textbf{Remark~6:}
We finally mention the readout delays of events.
In the current experimental setup, 
we focused on simple and practical phase objects represented by low-order Zernike polynomials. 
For these objects, the RMSE values in optical experiments (Fig.~\ref{fig:opt_p}) were comparable to those in computer simulations without the delay (Fig.~\ref{fig:sim_p}). 
Furthermore, events were accumulated over a sufficiently long time (1~second).
Based on these results, 
we believe that the effect of delay is nearly negligible for practical phase objects.
However, when the object becomes complex, e.g., a superposition of high-order Zernike polynomials, 
the delay probably becomes problematic and affects the quality of phase retrieval. 
We would like to address this issue in future work.
\vspace{3mm}

\section{Conclusion}
\label{s6}
We proposed a phase retrieval method using an event-based vision sensor (EVS)~\cite{Lichtsteiner:08}.
We showed that a wide dynamic range characteristic of the EVS allows us to rapidly and stably measure the axial intensity derivative.
We also demonstrated, both theoretically and numerically, that the phase distribution can be recovered using the \textit{transport of event equation (TEE)} and the axial intensity derivative (defocus events).
We believe that our study will contribute to advancing adaptive optics.

\bibliographystyle{IEEEtran}
\bibliography{main}

\begin{thebibliography}{10}
\providecommand{\url}[1]{#1}
\csname url@samestyle\endcsname
\providecommand{\newblock}{\relax}
\providecommand{\bibinfo}[2]{#2}
\providecommand{\BIBentrySTDinterwordspacing}{\spaceskip=0pt\relax}
\providecommand{\BIBentryALTinterwordstretchfactor}{4}
\providecommand{\BIBentryALTinterwordspacing}{\spaceskip=\fontdimen2\font plus
\BIBentryALTinterwordstretchfactor\fontdimen3\font minus \fontdimen4\font\relax}
\providecommand{\BIBforeignlanguage}[2]{{%
\expandafter\ifx\csname l@#1\endcsname\relax
\typeout{** WARNING: IEEEtran.bst: No hyphenation pattern has been}%
\typeout{** loaded for the language `#1'. Using the pattern for}%
\typeout{** the default language instead.}%
\else
\language=\csname l@#1\endcsname
\fi
#2}}
\providecommand{\BIBdecl}{\relax}
\BIBdecl

\bibitem{Shechtman:15}
Y.~Shechtman, Y.~C. Eldar, O.~Cohen, H.~N. Chapman, J.~Miao, and M.~Segev, ``Phase retrieval with application to optical imaging: A contemporary overview,'' \emph{IEEE Signal Processing Magazine}, vol.~32, no.~3, pp. 87--109, 2015.

\bibitem{Tyson:22}
R.~K. Tyson and B.~W. Frazier, \emph{Principles of Adaptive Optics}, 5th~ed.\hskip 1em plus 0.5em minus 0.4em\relax CRC Press, 2022.

\bibitem{Burge:76}
R.~E. Burge, M.~A. Fiddy, A.~H. Greenaway, and G.~Ross, ``The phase problem,'' in \emph{Proceedings of the Royal Society of London. Series A, Mathematical and Physical Sciences}, vol. 350, 1976, p. 191–212.

\bibitem{Wood:81}
J.~W. Wood, M.~A. Fiddy, and R.~E. Burge, ``Phase retrieval using two intensity measurements in the complex plane,'' \emph{Optics Letters}, vol.~6, no.~11, pp. 514--516, 1981.

\bibitem{Teague:83}
M.~R. Teague, ``Deterministic phase retrieval: A {Green’s} function solution,'' \emph{Journal of the Optical Society of America}, vol.~73, no.~11, pp. 1434--1441, 1983.

\bibitem{Streibl:84}
N.~Streibl, ``Phase imaging by the transport equation of intensity,'' \emph{Optics Communications}, vol.~49, no.~1, pp. 6--10, 1984.

\bibitem{Gerchberg:72}
R.~W. Gerchberg and W.~O. Saxton, ``A practical algorithm for the determination of phase from image and diffraction plane pictures,'' \emph{Optik}, vol.~35, pp. 237--246, 1972.

\bibitem{Fienup:82}
J.~R. Fienup, ``Phase retrieval algorithms: A comparison,'' \emph{Applied Optics}, vol.~21, no.~15, pp. 2758--2769, 1982.

\bibitem{Candes:15}
E.~Candes, X.~Li, and M.~Soltanolkotabi, ``Phase retrieval via {Wirtinger} flow: Theory and algorithms,'' \emph{IEEE Transactions on Information Theory}, vol.~61, no.~4, pp. 1985--2007, 2015.

\bibitem{Goldstein:18}
T.~Goldstein and C.~Studer, ``{PhaseMax}: Convex phase retrieval via basis pursuit,'' \emph{IEEE Transactions on Information Theory}, vol.~64, no.~4, pp. 2675--2689, 2018.

\bibitem{Allen:01}
L.~J. Allen and M.~Oxley, ``Phase retrieval from series of images obtained by defocus variation,'' \emph{Optics Communications}, vol. 199, no. 1-4, pp. 65--75, 2001.

\bibitem{Guirao:03}
A.~Guirao and D.~R. Williams, ``A method to predict refractive errors from wave aberration data,'' \emph{Optometry and Vision Science}, vol.~80, no.~1, pp. 36--42, 2003.

\bibitem{Xu:15}
R.~Xu, A.~Bradley, N.~L. Gil, and L.~N. Thibos, ``Modelling the effects of secondary spherical aberration on refractive error, image quality and depth of focus,'' \emph{Ophthalmic and Physiological Optics}, vol.~35, no.~1, pp. 28--38, 2015.

\bibitem{Shirai:09}
T.~Shirai, K.~Takeno, H.~Arimoto, and H.~Furukawa, ``Adaptive optics with a liquid-crystal-on-silicon spatial light modulator and its behavior in retinal imaging,'' \emph{Japanese Journal of Applied Physics}, vol.~48, no.~7R, p. 070213, 2009.

\bibitem{Waller:10a}
L.~Waller, S.~S. Kou, C.~J.~R. Sheppard, and G.~Barbastathis, ``Single-shot {TIE} using polarization multiplexing for quantitative phase imaging,'' \emph{Optics and Lasers in Engineering}, vol. 151, no.~15, p. 106912, 2010.

\bibitem{Waller:10b}
L.~Waller, Y.~Luo, S.~Y. Yang, and G.~Barbastathis, ``Transport of intensity phase imaging in a volume holographic microscope,'' \emph{Optics Letters}, vol.~35, no.~17, pp. 2961--2963, 2010.

\bibitem{Angel:21}
J.~A. Picazo-Bueno and V.~Mic\'{o}, ``Optical module for single-shot quantitative phase imaging based on the transport of intensity equation with field of view multiplexing,'' \emph{Optics Express}, vol.~29, no.~24, pp. 39\,904--39\,919, 2021.

\bibitem{Hai:22}
N.~Hai, R.~Kumar, and J.~Rosen, ``Phase from chromatic aberrations,'' \emph{Optics Express}, vol.~18, no.~22, pp. 22\,817--22\,825, 2010.

\bibitem{Kingslake:10}
R.~Kingslake and R.~B. Johnson, \emph{Lens Design Fundamentals}, 2nd~ed.\hskip 1em plus 0.5em minus 0.4em\relax Academic Press, 2010.

\bibitem{Lichtsteiner:08}
P.~Lichtsteiner, C.~Posch, and T.~Delbruck, ``A {$128\times 128$} {$120$} {dB} 15 \textmu s latency asynchronous temporal contrast vision sensor,'' \emph{IEEE Journal of Solid-State Circuits}, vol.~43, no.~2, pp. 566--576, 2008.

\bibitem{Garner:98}
L.~F. Garner, C.~S. Ooi, and G.~Smith, ``Refractive index of the crystalline lens in young and aged eyes,'' \emph{Clinical and Experimental Optometry}, vol.~81, no.~4, pp. 145--150, 1998.

\bibitem{Wiener:49}
N.~Wiener, \emph{Extrapolation, Interpolation, and Smoothing of Stationary Time Series: With Engineering Applications}.\hskip 1em plus 0.5em minus 0.4em\relax MIT Press, 1949.

\bibitem{Foi:08}
A.~Foi, M.~Trimeche, V.~Katkovnik, and K.~Egiazarian, ``Practical {Poissonian-Gaussian} noise modeling and fitting for single-image raw-data,'' \emph{IEEE Transactions on Image Processing}, vol.~17, no.~10, pp. 1737--1754, 2008.

\bibitem{Luisier:11}
F.~Luisier, T.~Blu, and M.~Unser, ``Image denoising in mixed {Poisson-Gaussian} noise,'' \emph{IEEE Transactions on Image Processing}, vol.~20, no.~3, pp. 696--708, 2011.

\bibitem{Peng:20}
Y.~Peng, S.~Choi, N.~Padmanaban, and G.~Wetzstein, ``Neural holography with camera-in-the-loop training,'' \emph{ACM Transactions on Graphics}, vol.~39, no.~6, pp. 1--14, 2020.

\bibitem{Yamamoto:21}
K.Yamamoto, I.~Suzuki, K.~Namikawa, K.~Sato, and Y.~Ochiai, ``Interactive eye aberration correction for holographic near-eye display,'' in \emph{Proceedings of the Augmented Humans International Conference}, 2021.

\bibitem{Hu:21}
Y.~Hu, S.~C. Liu, and T.~Delbruck, ``v2e: From video frames to realistic {DVS} events,'' in \emph{Proceedings of IEEE/CVF Conference on Computer Vision and Pattern Recognition Workshops}, 2021.

\bibitem{Foi:22}
A.~Foi, M.~Trimeche, V.~Katkovnik, and K.~Egiazarian, ``Zernike polynomials and their applications,'' \emph{Journal of Optics}, vol.~24, p. 123001, 2022.

\bibitem{Kelly:14}
D.~P. Kelly, ``Numerical calculation of the {Fresnel} transform,'' \emph{Journal of the Optical Society of America A}, vol.~31, no.~4, pp. 755--764, 2014.

\end{thebibliography}

\end{document}